# Layer-Selective Interlayer Exciton-Polarons in a *2H*-$MoS_2$ bilayer

*Seongwon Gim,[1] Dianne Lee,[1] Hyunjun Jo,[1] Yongju Jeong,[1] Kenji Watanabe,[2] Takashi Taniguchi,[3] and Hyeonjun Baek,[1,*]*

[1]Department of Physics, Sogang University, Seoul 04107, South Korea
[2]Research Center for Electronic and Optical Materials, National Institute for Materials Science, 1-1 Namiki, Tsukuba 305-0044, Japan
[3]Research Center for Materials Nanoarchitectonics, National Institute for Materials Science, 1-1 Namiki, Tsukuba 305-0044, Japan

*Corresponding author: HJBaek@sogang.ac.kr

## ABSTRACT

Exciton-Fermi polarons in two-dimensional (2D) semiconductors provide a platform for exploring quantum many-body interactions between bosons and fermions. While exciton-polaron physics has been extensively investigated in monolayer transition-metal dichalcogenides, the role of the layer degree of freedom in bilayer systems remains largely unexplored. Here, we investigate interlayer exciton (IX) polarons in electrically gated 2*H*-stacked bilayer $MoS_2$ using reflectance spectroscopy. Through layer-selective electron doping, we reveal a counterintuitive dependence of the polaron response on the spatial configuration of the interacting particles. When excess electrons occupy the electron layer of an IX, the attractive polaron branch becomes dominant, whereas electrons occupying the hole layer give rise to a predominantly repulsive polaron response, despite the stronger Coulomb attraction expected in the latter configuration. Calculations based on the ladder T-matrix approximation attribute this contrasting spectral-weight redistribution to the interplay between the binding energy of the underlying IX-electron bound state and the Fermi energy. Furthermore, the attractive polaron exhibits a pronounced energy splitting under hole-side doping, suggesting enhanced exchange interactions within the underlying IX-electron bound states. Our findings establish the layer degree of freedom as a control parameter for exciton-Fermi sea interactions, opening opportunities to explore spatially engineered many-body interactions in 2D quantum systems.

Two-dimensional (2D) transition-metal dichalcogenides (TMDs) have emerged as a particularly versatile platform for investigating exciton-polaron physics [1-6]. Quantum confinement and reduced dielectric screening give rise to tightly bound excitons with binding energies of several hundred meV[7,8], while the atomically thin nature of these materials enables precise electrostatic control of the density of itinerant charge carriers. The coexistence of robust excitons and electrically tuneable Fermi seas, therefore, allows systematic investigation of exciton–carrier interactions over a wide range of carrier densities[9]. The resulting optical response is characterised by the emergence of attractive and repulsive polaron (AP and RP) branches and the redistribution of oscillator strength between them with increasing carrier density.

Beyond monolayer TMDs, TMD-based bilayer and multilayer systems have emerged as a versatile platform for exploring correlated quantum phenomena, hosting a variety of emergent quantum phases, including exciton condensates[10,11], electronic and excitonic Mott insulating states[12-17], Wigner crystals[18,19], excitonic insulators[20-23], and Chern insulators[24,25]. Unlike monolayer systems, bilayers introduce an additional degree of freedom, the layer index, which enables control over the spatial configuration of interacting particles[26,27]. In particular, the spatial separation of the electron and hole in an interlayer exciton (IX) provides two distinct interaction configurations depending on which layer is occupied by excess carriers[28,29]. From a simple electrostatic perspective, excess electrons residing in the hole layer of an IX would be expected to experience stronger attraction, whereas those residing in the electron layer would experience stronger repulsion. Furthermore, the relative layer configuration modifies the spatial overlap and accessible spin-valley configurations of the interacting particles, potentially giving rise to distinct electron-electron and electron-hole exchange interactions. How these distinct microscopic interactions manifest themselves in the many-body optical response, however, remains unclear. Though previous layer-resolved studies have employed intralayer excitons interacting with carriers in an adjacent layer[30,31], polarons formed by the IX itself have been difficult to access. Owing to the weak oscillator strength of conventional IXs, previous investigations of charged IX complexes have largely relied on photoluminescence (PL) spectroscopy[32-37]. However, PL spectroscopy provides limited access to the repulsive polaron branch and does not directly quantify the transfer of oscillator strength between the AP and RP branches, both of which are key spectroscopic signatures of exciton-polaron physics[38]. Reflectance spectroscopy of optically active IXs therefore offers a means of exploring these layer-dependent many-body interactions.

Here, we investigate IX-Fermi polarons in a 2*H*-stacked bilayer $MoS_2$ using gate-dependent reflectance spectroscopy. The coupling of IXs to intralayer B excitons ($X_B$) provides sufficient oscillator strength to directly probe their many-body interactions with itinerant electrons[39-41]. By selectively controlling the electron population in individual layers, we reveal a counterintuitive layer-dependent polaron response. The AP branch becomes dominant when excess electrons occupy the electron layer of an IX, whereas the RP branch retains a large fraction of the oscillator strength when excess electrons occupy the hole layer. These contrasting behaviours are reproduced by a Green's-function formalism within the ladder *T*-matrix approximation, which identifies the binding energy of the underlying IX-electron bound state relative to the Fermi energy, $E_F$, as the key parameter governing the redistribution of oscillator strength[3]. Moreover, electric-field-dependent spectroscopy reveals a pronounced splitting of the AP under hole-side doping, which we attribute to enhanced exchange interactions enabled by the distinct spatial and spin-valley configurations of the interacting particles. Our findings establish the layer degree of freedom as an additional means of controlling both the spectral weight distribution and fine structure of exciton-Fermi polarons in two-dimensional Bose-Fermi mixtures.

## Results

### Interlayer exciton-polaron formation

We investigate IX polarons in an hBN-encapsulated 2*H*-$MoS_2$ bilayer with top and bottom graphite gates (Fig. 1a). The dual-gate geometry enables independent control of the carrier density, $n$, and the out-of-plane electric field, $F_z$. The bottom and top $MoS_2$ layers are denoted as $L_1$ and $L_2$, respectively.

The bilayer hosts two IX configurations with opposite out-of-plane electric dipole moments, as shown in the bottom panel of Fig. 1a. Coupling to intralayer $X_B$ endows these IXs with sufficient oscillator strength for their detection in reflectance spectroscopy.

As in the case of intralayer excitons, charged complexes of IXs can form upon electrostatic doping. The spatial separation of the electron and hole in an IX, however, introduces distinct interaction channels depending on the layer occupied by the excess carriers. Figures 1b, c schematically illustrate IXs with a downward- and upward-oriented electric dipole ($IX_\downarrow$ and $IX_\uparrow$) interacting with an electron Fermi sea confined to $L_2$. The dressing of the IX is represented by particle-hole excitations of the Fermi sea. Depending on the relative valley configurations of the IX and the interacting electrons, the dressing can occur through either intravalley or intervalley scattering channels. For $IX_\downarrow$ (Fig. 1b), the constituent electron resides in the same layer as the electron Fermi sea. Consequently, the electron-electron repulsion is intralayer, whereas the attractive interaction between the excess electron and the IX hole is interlayer. Conversely, for $IX_\uparrow$ (Fig. 1c), the spatial configuration is reversed. The IX hole resides in the electron-doped layer, whereas the IX electron occupies the opposite layer. Thus, the electron-hole attraction is intralayer, while the electron-electron repulsion is interlayer. These distinct spatial configurations provide a means of controlling IX-electron interactions and the resulting IX-polaron formation through layer-selective doping. In addition to modifying the balance between attractive and repulsive Coulomb interactions, these distinct spatial configurations modify the exchange interactions and accessible spin-valley configurations of the underlying charged complexes. Layer-selective doping therefore provides a means of controlling the microscopic interaction channels involved in IX-polaron formation.

Figure 1d presents the second derivative of the reflectance contrast with respect to photon energy, $\frac{d^2(\Delta R/R_0)}{dE^2}$, as a function of carrier density. At charge neutrality, the intralayer A-exciton ($X_A$) appears at 1.94 eV, and the IX is observed at 2.01 eV, accompanied by two additional resonances associated with its 2s and 3s Rydberg states at 2.09 and 2.10 eV, respectively. These spectral assignments are further supported by the electric-field-dependent measurements discussed below. Upon electron or hole doping, the $X_A$ resonance evolve into AP and RP branches (see Supplementary Fig. S1b, c for measurements from another device). From the energy separation between the neutral exciton and the AP resonance near the onset of doping, the corresponding trion binding energies are estimated to be 21 and 14 meV for electron and hole doping, respectively. Interestingly, polaron features also emerge from the IX resonance, exhibiting qualitatively different spectral responses to electron and hole doping. In the electron-doped regime, two distinct resonances are identified as the AP and RP branches of the IX as shown in the enlarged spectral map in Fig. 1e. The energy separation between the AP near the onset of electron doping and the neutral IX yields an estimated IX-electron binding energy of 6 meV (Fig. 1f), considerably smaller than that of the intralayer trion associated with $X_A$[35-37]. In contrast, under hole doping, only the RP branch of the IX is observed, with no clearly resolved AP resonance (see Supplementary Fig. S1b, c).

## Layer-selective doping and asymmetric polaron formation

Although the interaction between an IX and excess carriers is expected to depend on their relative layer configurations, the near degeneracy of the electronic states in layers $L_1$ and $L_2$ allows electrons to populate both layers, making it difficult to distinguish the individual interaction channels. To isolate these layer-dependent interactions, we apply a finite out-of-plane electric field to lift the layer degeneracy and enable preferential carrier occupation of individual layers. Figure 2a shows the evolution of $\frac{d^2(\Delta R/R_0)}{dE^2}$ as a function of gate voltage $V_G$ at a fixed electric field of $F_z = 33$ mV nm$^{-1}$. The carrier density is varied by sweeping $V_G$ while maintaining a constant voltage offset between the top and bottom gates, with $V_T = V_G + 2\ V$ and $V_B = V_G - 2\ V$ (see Supplementary Section S2 for additional measurements with different voltage offsets). First, the evolution of the $X_A$ resonances reveals sequential electron filling of $L_2$ and $L_1$. At the onset of electron doping, electrons preferentially occupy $L_2$, giving rise to the AP and RP branches associated with $X_A$ in this layer. Meanwhile, the $X_A$ resonance in $L_1$ undergoes a redshift of 7 meV without developing a resolved polaron splitting, consistent with the preferential occupation of the opposite layer. With further electron doping, $L_1$ also becomes populated, leading to the emergence of its own AP and RP branches. The RP resonance associated with $X_A$ in $L_1$ exhibits a smaller density-dependent energy shift than its counterpart in $L_2$, reflecting the distribution of electrons between the two layers (see Supplementary Section S3 for additional measurements). In contrast, no comparable sequential filling behaviour is observed under hole doping. This difference may be attributed to the distinct layer characteristics of the relevant conduction and valence band states. Whereas conduction band states can exhibit a preferential layer character under an applied electric field, the valence band states near the Γ point, which are expected to accommodate the doped holes, are delocalised across the two layers[42,43]. Consequently, hole doping does not exhibit the same pronounced layer-selective behaviour as electron doping.

Notably, a contrasting response emerges in the IX-polaron spectra depending on the orientation of the IX electric dipole, as shown in the enlarged spectral map in Fig. 2b. For the $IX_\downarrow$, the AP branch ($IX_\downarrow^{AP-}$) becomes substantially stronger than the RP branch ($IX_\downarrow^{RP-}$) with increasing electron density. In contrast, the $IX_\uparrow$ spectrum is dominated by the RP branch ($IX_\uparrow^{RP-}$), with no discernible AP resonance in this spectral region. To quantitatively compare these responses, the spectral weights of the polaron resonances, normalised to those of their respective parent neutral IXs, are plotted as a function of carrier density in Fig. 2d. The normalised spectral weight of $IX_\downarrow^{RP-}$ decreases markedly from 0.89 to 0.30, whereas that of $IX_\uparrow^{RP-}$ decreases only slightly from 0.79 to 0.68 as the electron density increases from $0.4\times10^{12}$ to $0.9\times10^{12}$ cm$^{-2}$. The persistence of the dominant $IX_\uparrow^{RP-}$ suggests that only a small fraction of its original oscillator strength is transferred to the AP. The contrasting redistribution of oscillator strength between the two IX configurations demonstrates that the optical response of IX polarons is governed not only by carrier density but also by the relative layer configuration of the interacting particles.

The origin of this asymmetry can be understood by considering the spatial configurations of the underlying IX-electron bound states, as illustrated in Fig. 1b, c. For $IX_\downarrow$, an additional electron can bind to the IX to form a negatively charged complex consisting of two electrons in one layer and a hole in the opposite layer. In this configuration, the attractive interaction between the additional electron and the IX hole is interlayer, whereas the repulsion between the two electrons is intralayer. The resulting balance of Coulomb interactions is consistent with the small binding energy determined experimentally. Consequently, the trion-like bound state lies close in energy to the parent IX resonance, facilitating the transfer of oscillator strength to the AP branch, as discussed below within the Green's-function formalism. For $IX_{\uparrow}$, the spatial configuration is reversed (Fig. 1c). The additional electron experiences intralayer attraction to the IX hole and interlayer repulsion from the IX electron, favouring the formation of a more deeply bound state. This configuration contains an intralayer electron-hole pair accompanied by an additional electron in the opposite layer, and the corresponding optical transition energy can approach that of the intralayer $X_A$. The bound state thus lies well below the parent IX resonance. Despite the stronger microscopic attraction in this configuration, the optical response remains dominated by the RP, with only a weak contribution from the AP. As demonstrated below, this counterintuitive behaviour arises from the large binding energy of the underlying IX-electron complex relative to $E_F$, which limits the transfer of oscillator strength to the AP branch at experimentally accessible carrier densities.

Meanwhile, both $IX_{\downarrow}$ and $IX_{\uparrow}$ exhibit predominantly RP-like responses under hole doping, with no clearly resolved AP branches in the density-dependent spectra (see Supplementary Fig. S1b, c). This behaviour may be related to the delocalised nature of the valence band states near the Γ point. Since the doped-hole wavefunctions extend across both $MoS_2$ layers, their interaction with the IX is expected to be less sensitive to the orientation of the IX electric dipole moment. The resulting interaction configurations may favour deeply bound charged complexes, analogous to those formed when an IX interacts with an electron Fermi sea in its hole layer. Such a scenario could explain the weak AP spectral weight and the persistence of the RP under hole doping.

**Electric-field-dependent evolution of interlayer exciton-polarons**

To further investigate the layer-dependent IX-polaron response, we examine the evolution of the excitonic resonances with the out-of-plane electric field at selected carrier densities. Figures 3a-c present the electric-field-dependent spectra at charge neutrality and at electron densities of $0.9\times10^{12}$ and $2.2\times10^{12}$ cm$^{-2}$, respectively. At charge neutrality, both the IX and its 2s and 3s Rydberg states exhibit linear Stark shifts, reflecting their permanent out-of-plane electric dipole moments and supporting their spectral assignments. At finite electron densities, varying the electric field modifies the distribution of electrons between the two layers, as evidenced by the evolution of $X_A$ and its associated polaron branches. Specifically, as the electric field increases from 0 to 17 mV nm$^{-1}$ in Fig. 3c, electrons are depleted from $L_1$ and accumulate in $L_2$. Accordingly, the $X_A$ resonance in $L_1$ evolves from a polaronic feature towards the neutral exciton resonance, whereas its counterpart in $L_2$ develops the characteristic response of an electron-doped exciton.

The IX spectra exhibit a qualitatively different evolution. In contrast to the neutral IX resonances, which become degenerate near zero electric field, the IX-derived resonances remain split at $F_z = 0$ owing to the formation of energetically distinct AP and RP branches. The lower- and higher-energy resonances are assigned to the AP and RP branches, respectively. Notably, an additional weak spectral feature emerges upon electron doping and exhibits a nearly linear Stark shift, as indicated by the yellow arrows in Fig. 3b, c. Its electric field dependence reveals its dipolar character, while extrapolation towards zero electric field places its energy close to that of $X_A$. We assign this feature to the AP of $IX_\uparrow$ ($IX_\uparrow^{AP-}$), associated with the deeply bound IX-electron configuration discussed above. Although this AP is difficult to distinguish in the density-dependent spectra of Fig. 2a owing to its weak oscillator strength, its characteristic Stark shift enables its identification in the electric-field-dependent measurements. The observation of this weak, low-energy AP further supports the connection between the binding energy of the underlying IX-electron complex and the redistribution of oscillator strength between the polaron branches.

Remarkably, the AP of $IX_\uparrow$ exhibits a doublet with an energy separation of 12 meV, whereas no discernible splitting is observed in the AP of $IX_\downarrow$ or in the corresponding RP branches. This contrasting behaviour suggests that exchange interactions within the underlying IX-electron bound states are sensitive to the relative layer configuration of the interacting particles. For electron-side doping, the two electrons reside in the same layer, imposing restrictions on their accessible spin-valley configurations through the Pauli exclusion principle. In particular, configurations involving electrons with parallel spins require different valley indices within the configurations considered here, giving rise to intervalley electron-electron exchange, as illustrated in Fig. 1b. However, the weak binding and associated extended spatial distribution of this charged complex are expected to reduce the electron-electron overlap and suppress the short-range contribution to the exchange interaction. In contrast, for hole-side doping, the spatial separation of the two electrons into different layers relaxes these spin-valley restrictions, permitting configurations in which the electrons have identical spin and valley indices. More importantly, the additional electron resides in the same layer as the IX hole, enabling spin- and valley-dependent intralayer electron-hole exchange interactions (Fig. 1c). The stronger binding and enhanced spatial overlap between the additional electron and the IX hole are expected to increase the electron-hole exchange energy, potentially accounting for the pronounced AP splitting observed under hole-side doping (see Supplementary Section S4 for further discussion). Notably, the observed splitting of 12 meV is comparable to the electron-hole exchange energy previously reported for monolayer $MoS_2$[44], supporting our interpretation. Interestingly, despite the pronounced AP splitting, no corresponding splitting is resolved in the RP branch. This behaviour is consistent with two energetically distinct bound-state scattering channels coupled to a common optically active bare IX state. Within such a picture, the two bound-state channels give rise to distinct AP resonances, whereas their contributions to the exciton self-energy collectively renormalise the RP branch without necessarily producing a resolvable splitting. These observations suggest that the layer degree of freedom controls not only the overall strength of IX-electron interactions but also the exchange interactions that determine the fine structure of the resulting exciton polarons.

**Simulation of interlayer exciton-polaron spectra**

To elucidate the origin of the layer-dependent IX-polaron response, we calculate the exciton spectral functions using a Green's-function formalism within the ladder T-matrix approximation[2,3]. Exciton-electron interactions are incorporated through a finite-density exciton self-energy that accounts for the binding energies of the corresponding IX-electron bound states. To spectrally distinguish the polaron branches associated with $IX_{\downarrow}$ and $IX_{\uparrow}$, the calculations are performed under a finite electric field, corresponding to the experimental configuration in Fig. 2a (see Supplementary Section S5 for details). Figure 4a presents the calculated IX-polaron spectra as a function of carrier density. With increasing electron density, AP and RP branches develop from the neutral IX resonances, exhibiting distinct spectral-weight redistributions for the two IX configurations. For $IX_{\downarrow}$, oscillator strength is efficiently transferred to the AP, accompanied by the suppression of the RP. In contrast, for $IX_{\uparrow}$, the RP retains a large fraction of the oscillator strength, while the two AP resonances emerge at lower energies. The two AP resonances arise from the inclusion of two bound-state scattering channels with different binding energies. The calculations thus capture the essential features of the experimentally observed layer-dependent polaron response. Figures 4b, c show the density-dependent spectral weights of the AP and RP branches, normalised to those of the corresponding neutral IXs, as extracted from the calculated spectra in Fig. 4a. For a quantitative comparison, the experimental spectral weights of the dominant polaron branches for the two IX configurations, shown in Fig. 2d, are plotted together with the calculated values in Fig. 4d. To account for uncertainties in the spin-valley degeneracy used to estimate $E_F$ and in the experimental calibration of the carrier density, a constant scaling factor of 1.23 is applied to the carrier density in the calculations ($n_{cal} = \alpha n_{exp}$) (see Supplementary Section S6 for the unscaled results). The calculated spectral weights are in good quantitative agreement with the experimental results, reproducing the dominance of the AP branch for $IX_{\downarrow}$ and the RP branch for $IX_{\uparrow}$.

The central distinction between the two IX configurations is the binding energy of the IX-electron bound state relative to $E_F$ of the excess electrons. For the electron-side-doped IX, the experimentally inferred binding energy is small, $E_{T,e}$ = 6 meV, such that $E_F$ becomes comparable to $E_{T,e}$ at experimentally accessible electron densities. For example, an electron density of approximately $10^{12}$ $cm^{-2}$ corresponds to an $E_F$ of 4–5 meV for the relevant conduction band parameters. The electron-side-doped IX therefore enters the polaron crossover regime, $E_F \sim E_{T,e}$, in which coupling to Fermi sea excitations leads to efficient redistribution of oscillator strength from the RP to the AP. In contrast, for the hole-side-doped IX, the corresponding bound state lies much farther below the parent IX resonance, with experimentally inferred binding energies of $E_{T,h} = 58$ and 70 meV, assigned to the intravalley and intervalley configurations, respectively. Consequently, the hole-side-doped IX remains in the regime $E_F \ll E_{T,h}$ for both scattering channels throughout the experimentally accessible density range. In the low-density limit of the exciton-polaron model[3], the AP spectral weight scales approximately as $\frac{E_F}{E_T}$. At a given electron density, a larger bound-state binding energy therefore results in a smaller transfer of oscillator strength to the AP. This explains why the hole-side-doped IX retains a dominant RP response despite the stronger attraction between the IX and the excess electrons. Conversely, the weakly bound state associated with the electron-side-doped IX gives rise to a dominant AP once the Fermi energy becomes comparable to its binding energy. These results reveal that the relative layer configuration of an IX and a Fermi sea governs not only the binding energy of the underlying exciton-electron complex but also the redistribution of oscillator strength between the resulting polaron branches. In particular, a stronger microscopic attraction does not necessarily lead to a stronger attractive polaron response. Rather, the optical manifestation of exciton-electron interactions is governed by the interplay between the bound-state binding energy and $E_F$, providing a means of controlling exciton-polaron formation through the layer degree of freedom.

## Discussion

In conclusion, we demonstrate layer-dependent IX-Fermi polaron formation in $2H$-stacked bilayer $MoS_2$. By combining layer-selective carrier doping with reflectance spectroscopy, we distinguish two interaction configurations in which the electron Fermi sea resides in either the electron or the hole layer of an IX. The former exhibits efficient transfer of oscillator strength to the attractive polaron branch, whereas the latter retains most of its oscillator strength in the repulsive polaron branch. Furthermore, electric-field-dependent spectroscopy reveals a weak attractive polaron doublet for hole-side-doped IX, with an energy splitting of 12 meV. The contrasting polaron responses are captured by a Green's-function formalism within the ladder T-matrix approximation, identifying the relative energy scales of exciton-electron binding and the Fermi energy as key factors governing the redistribution of oscillator strength. Beyond conventional carrier-density control, our work establishes the spatial configuration of interacting particles as an additional degree of freedom for tailoring quantum many-body interactions in layered semiconductors. This understanding also provides a basis for interpreting excitonic signatures of correlated quantum phases in TMD-based bilayers.

During the preparation of this manuscript, we became aware of an independent study by Duan et al.[45] reporting dipolar exciton polarons in bilayer $MoS_2$.

# Methods

## Fabrication of dual-gated devices

Bilayer $MoS_2$, hexagonal boron nitride (hBN), and graphite flakes were mechanically exfoliated onto 90-nm-thick $SiO_2$ substrates. The flakes were subsequently picked up and assembled onto a $SiO_2$/Si substrate with pre-patterned Cr/Au (5/10 nm) electrodes using a polycarbonate (PC)-coated polydimethylsiloxane (PDMS) stamp in an Ar-filled glove box. After stacking, the device was immersed in chloroform to dissolve the PC layer.

## Reflectance contrast spectroscopy

Low-temperature reflectance measurements were performed using a confocal microscopy system equipped with a 100x objective lens (NA = 0.90). A halogen lamp was used as the illumination source. The sample was mounted on a piezoelectric stage for nanometre-scale positioning inside a closed-cycle cryostat (Montana Instrument, NFEC-2024-12-301453) maintained at 4 K. The excitation spot was controlled using galvanometer mirrors. Then the reflected light was dispersed by a spectrometer with a focal length of 500 mm and detected using a liquid-nitrogen-cooled charge-coupled device (CCD). To enhance the signal-to-noise ratio, the measurement was repeated 20–50 times, and the resulting spectra were averaged. The reflectance contrast was defined as $\frac{\Delta R}{R_0} = \frac{R - R_0}{R_0}$, where $R$ and $R_0$ denote the reflection signals from the sample and reference regions, respectively. To enhance the visibility of the excitonic resonances, the first or second derivatives of reflectance contrast with respect to photon energy was calculated.

## Reflectance contrast fitting

To determine the resonance energies and linewidths, the reflectance contrast spectra were fitted using a dispersive Lorentzian function[46]

$$RC\,(E) = A\,cos\varphi \frac{\gamma/2}{(E - E_0)^2 + \gamma^2/4} + A\,sin\varphi \frac{E_0 - E}{(E - E_0)^2 + \gamma^2/4} + C,$$

where $A$ is the amplitude, $\gamma$ is the full width at half-maximum, $E$ is the photon energy, $E_0$ is the resonance energy, $\varphi$ is the phase parameter, and $C$ is a background offset.

**Calculation of the electric field and carrier density**

The out-of-plane electric field, $F_z$, was calculated as $F_z = \frac{\varepsilon_{hBN}}{\varepsilon_{MoS_2}} \frac{V_T - V_B}{d_T + d_B}$, where $V_T$ and $V_B$ are the top- and bottom-gate voltages relative to the $MoS_2$ layer, and $d_T$ and $d_B$ are the thicknesses of the top and bottom hBN layers, both estimated to be 26 nm from optical contrast measurements. The relative permittivities of hBN and $MoS_2$ were taken to be $\varepsilon_{hBN} \approx 3$[12] and $\varepsilon_{MoS_2} \approx 7$[47]. The nominal gate-induced carrier density, $n$, was estimated using a parallel-plate capacitor model, $n = \frac{\varepsilon_0 \varepsilon_{hBN}}{e}\left(\frac{V_T}{d_T} + \frac{V_B}{d_B}\right)$, where $\varepsilon_0$ is the vacuum permittivity and $e$ is the elementary charge.

**Data availability**

Data described in this paper and presented in the Supplementary Information are available online.

**Acknowledgements**

This research is supported by the POSCO Science Fellowship of the POSCO TJ Park Foundation and the National Research Foundation (NRF) of Korea and Korea Basic Science Institute (National Research Facilities and Equipment Center) grants funded by the Korean Ministry of Science and ICT (Grant Nos. RS-2023-00220471, RS-2024-00404384, RS-2025-24683194, RS-2025-14532973, and RS-2025-25463680).

**Author contributions**

S.G. and D.L. fabricated the devices. S.G. performed the optical measurements and processed all experimental data. D.L. performed the remaining measurements with assistance from Y.J. H.J. provided technical support for the optical setup. K.W. and T.T. supplied the hexagonal boron nitride crystals. S.G. and H.B. analysed the data, performed the numerical simulations and wrote the manuscript. H.B. supervised the project.

**Competing interests**

The authors declare no competing interests.

## Figures

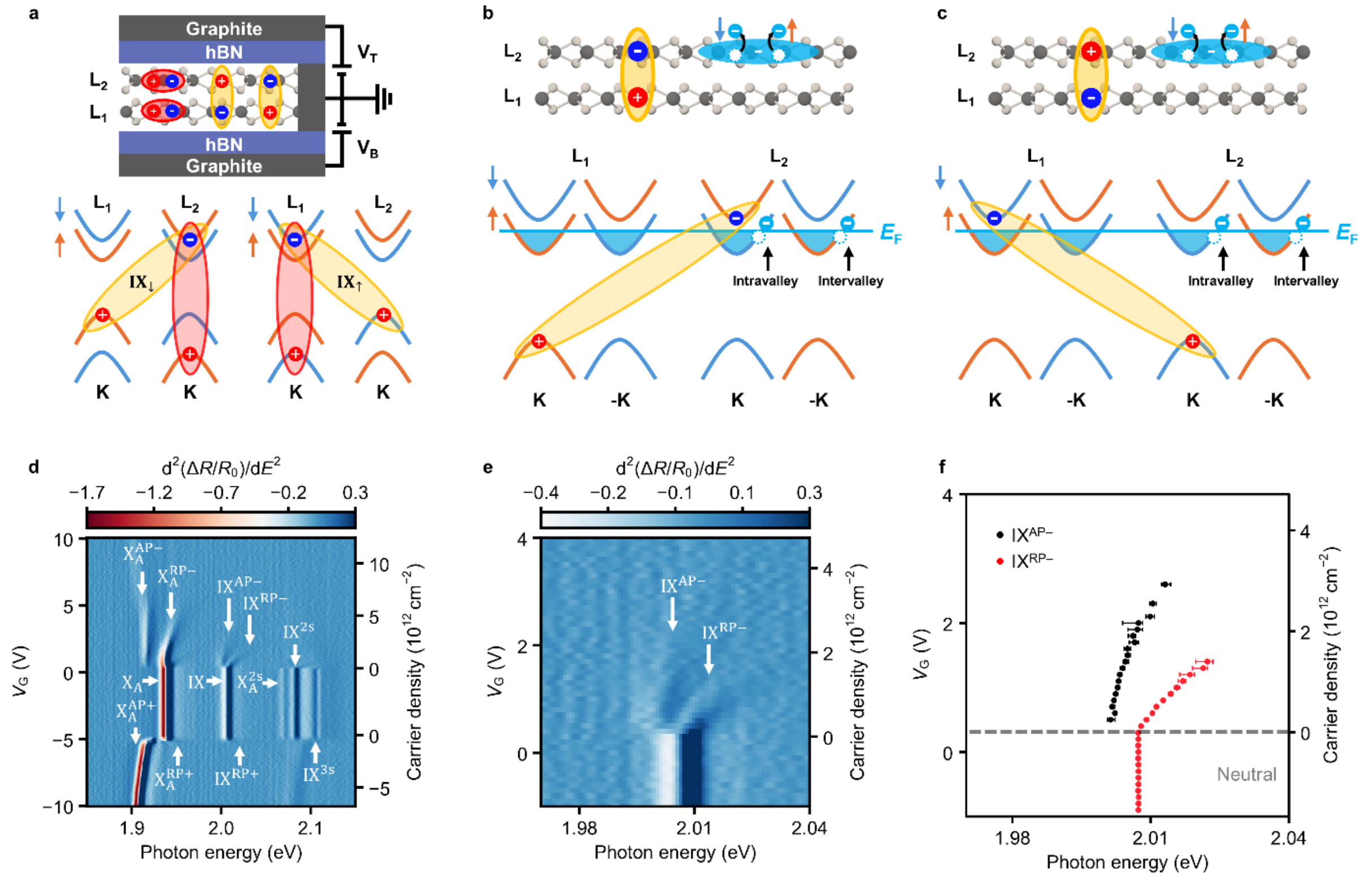

**Figure 1. Device structure and concept of IX-polaron formation.** **a** Schematic of the dual-gated device structure and IX configurations in a *2H*-$MoS_2$ bilayer. $L_1$ and $L_2$ denote the bottom and top $MoS_2$ layers, respectively. **b, c** Real-space charge configurations (top) and corresponding band diagrams (bottom) of $IX_\downarrow$ (**b**) and $IX_\uparrow$ (**c**) interacting with an electron Fermi sea in $L_2$. Only IX configurations associated with the K valley are displayed. Filled cyan circles and dashed open cyan circles represent particle-hole excitations of the Fermi sea. The horizontal cyan lines indicate $E_F$. **d** Spectral map of $d^2(\Delta R/R_0)/dE^2$ as a function of carrier density. **e, f** Enlarged view **(e)** and fitted resonance energies **(f)** of the IX spectral region in **d**. Black points indicate $IX^{AP-}$ and red points trace the neutral IX into $IX^{RP-}$.

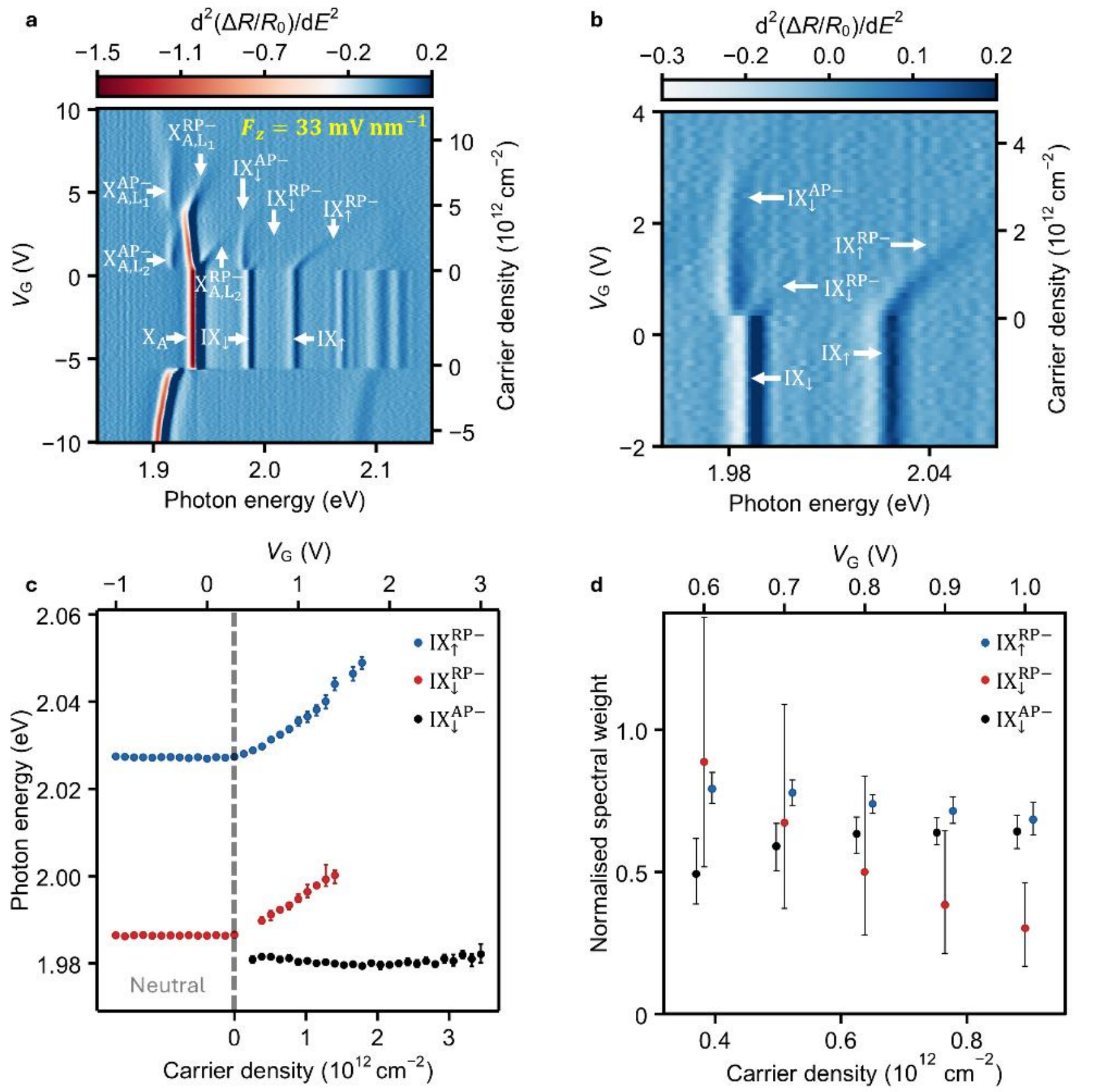

**Figure 2. Doping-dependent IX-polaron spectra under a fixed electric field. a** Spectral map of $d^2(\Delta R/R_0)/dE^2$ as a function of $V_G$ at a fixed electric field of $F_z = 33$ mV nm$^{-1}$. The carrier density is varied by sweeping $V_G$ while maintaining a constant voltage offset between the top and bottom gates, with $V_T = V_G + 2\,V$ and $V_B = V_G - 2\,V$. The right axis indicates the nominal carrier density, with positive and negative values corresponding to electron and hole doping, respectively. **b** Enlarged view of the IX spectral region in **a**. **c** Fitted resonance energies of the neutral IXs and their associated polaron branches. **d** Normalised spectral weights of $\mathrm{IX}_{\uparrow}^{\mathrm{RP}-}$, $\mathrm{IX}_{\downarrow}^{\mathrm{RP}-}$, and $\mathrm{IX}_{\downarrow}^{\mathrm{AP}-}$ as a function of carrier density. The spectral weight of each polaron branch is normalised to that of its corresponding parent neutral IX. For visual clarity, the experimental data for $\mathrm{IX}_{\uparrow}^{\mathrm{RP}-}$ ($\mathrm{IX}_{\downarrow}^{\mathrm{AP}-}$) are slightly displaced along the positive (negative) direction of the carrier density axis.

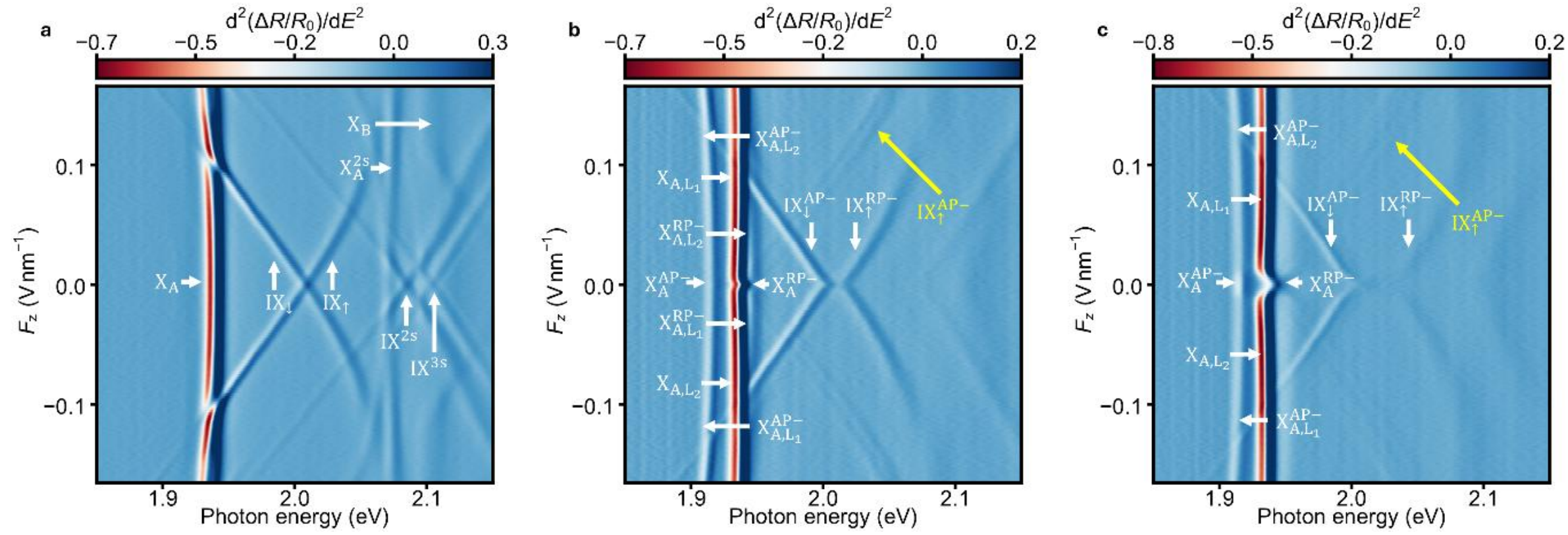

**Figure 3. Electric field dependence under fixed electron densities and emergence of $\mathrm{IX}_{\uparrow}^{\mathrm{AP}-}$. a-c** Spectral maps of $d^2(\Delta R/R_0)/dE^2$ as a function of $F_z$, measured at charge neutrality (**a**) and at electron densities of 0.9×10$^{12}$ (**b**) and 2.2×10$^{12}$ cm$^{-2}$ (**c**). The yellow arrows indicate $\mathrm{IX}_{\uparrow}^{\mathrm{AP}-}$ for $F_z >$ 0.

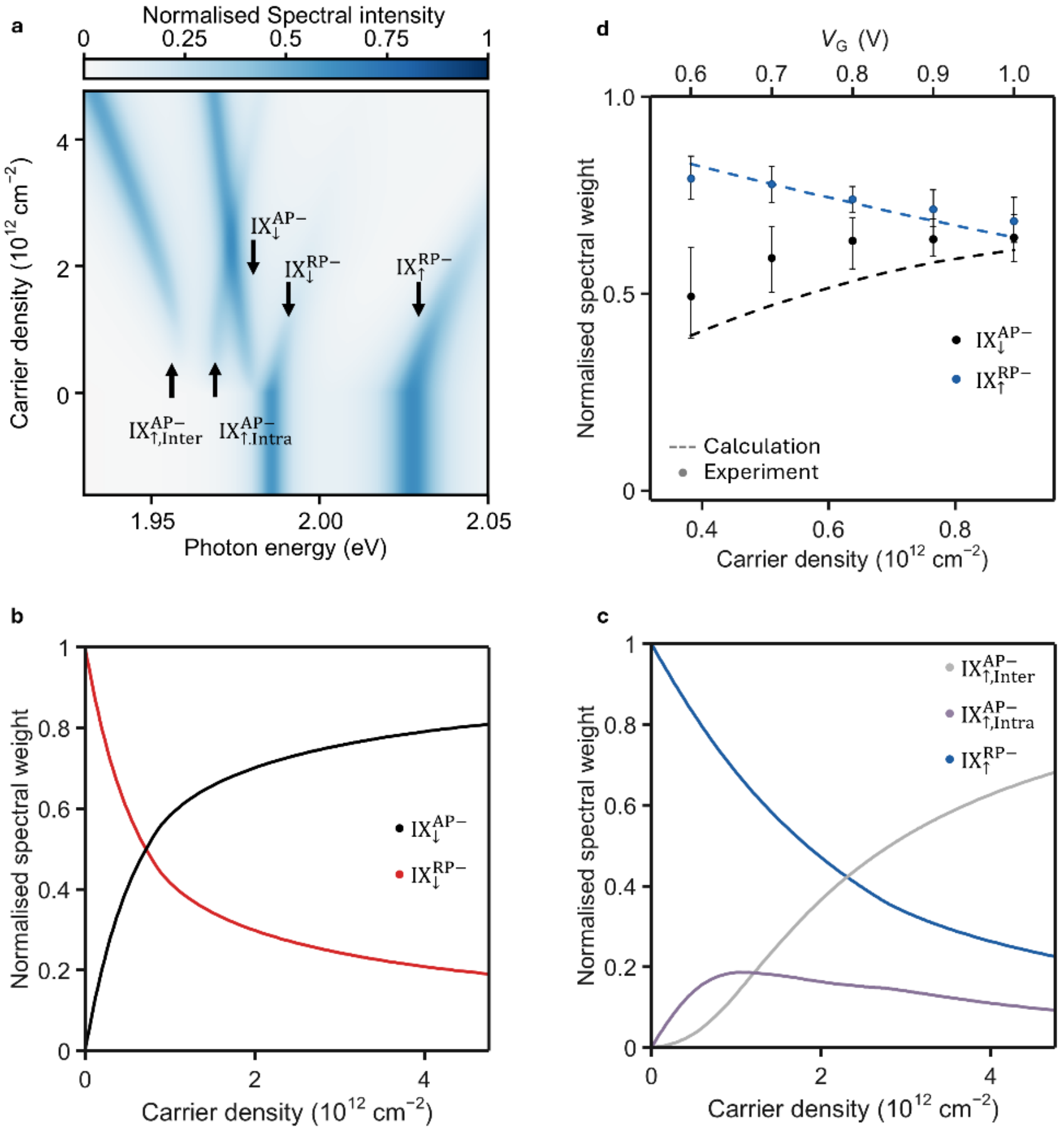

**Figure 4. Simulation of IX-polaron spectra. a** Calculated exciton spectral functions using a Green's-function formalism within the ladder T-matrix approximation. The binding energies of the IX-electron bound states are set to 6 meV for $IX_{\downarrow}$ and 58 and 70 meV for the intravalley and intervalley scattering channels of $IX_{\uparrow}$, respectively, based on experimentally obtained values. **b, c** Normalised spectral weights of the polaron branches associated with $IX_{\downarrow}$ (**b**) and $IX_{\uparrow}$ (**c**) as a function of electron density. The spectral weight of each polaron branch is normalised to that of its corresponding parent neutral IX. **d** Comparison between the experimentally measured and calculated spectral weights. A constant scaling factor of 1.23 is applied to the carrier density in the calculations.